\pdfoutput=1
\documentclass[aps,prl,twocolumn,10pt]{revtex4-1}
\usepackage[utf8]{inputenc}
\usepackage{fourier,amsmath,graphicx,microtype}
\usepackage[cal=boondoxo,frak=euler]{mathalfa}
\usepackage[colorlinks=true,citecolor=blue,urlcolor=blue]{hyperref}

\begin{document}

\title{Dynamic transport in a quantum wire driven by spin-orbit interaction}

\author{Yasha Gindikin}

\affiliation{Kotel'nikov Institute of Radio Engineering and Electronics,
Russian Academy of Sciences, Fryazino, Moscow District, 141190, Russia}

\begin{abstract}
We consider a gated one-dimensional (1D) quantum wire disturbed in a contactless manner by an alternating electric field produced by a tip of a scanning probe microscope. In this schematic 1D electrons are driven not by a pulling electric field but rather by a non-stationary spin-orbit interaction (SOI) created by the tip. We show that a charge current appears in the wire in the presence of the Rashba SOI produced by the gate net charge and image charges of 1D electrons induced on the gate (iSOI). The iSOI contributes to the charge susceptibility by breaking the spin-charge separation between the charge- and spin collective excitations, generated by the probe. The velocity of the excitations is strongly renormalized by SOI, which opens a way to fine-tune the charge and spin response of 1D electrons by changing the gate potential. One of the modes softens upon increasing the gate potential to enhance the current response as well as the power dissipated in the system.
\end{abstract}

\maketitle

\textit{Introduction}.---Today we are witnessing the burst of interest in the ballistic electron transport in quantum wires~\cite{Heedt2017,doi:10.1063/1.4977031,doi:10.1021/acs.nanolett.6b00414,0957-4484-26-21-215202,doi:10.1021/nl3035256}. For the last three decades the quantum wires formed by electrostatic gating of a high-mobility two-dimensional (2D) electron gas have been the favorite playground to study quantum many-body effects in one-dimensional (1D) electron systems~\cite{Clarke2016}, where a strongly correlated state known as the Tomonaga-Luttinger liquid emerges as a result of the electron-electron (e-e) interaction~\cite{0034-4885-58-9-002}. The dynamic transport experiments are the most subtle and precise methods to extract the many-body physics~\cite{Auslaender88}.

Currently of most interest are the group III-V semiconductor nanowires as they represent basic building blocks for the topological quantum computing~\cite{alicea2011non} and spintronics~\cite{bandyopadhyay2015introduction}. In particular, InAs and InSb nanowires are promising systems for the creation of helical states and as a host for Majorana fermions~\cite{0953-8984-25-23-233201,0034-4885-75-7-076501,doi:10.1146/annurev-conmatphys-030212-184337}. The fundamental reason behind these properties is the strong Rashba spin-orbit interaction (RSOI) in these materials~\cite{manchon2015new}.

Recently we have found that RSOI is created by the electric field of the image charges that electrons induce on a nearby gate~\cite{PhysRevB.95.045138}. A sufficiently strong image-potential-induced spin-orbit interaction (iSOI) leads to highly non-trivial effects such as the collective mode softening and subsequent loss of stability of the elementary excitations, which appear because of a positive feedback between the density of electrons and the iSOI magnitude.

By producing a spin-dependent contribution to the e-e interaction Hamiltonian of 1D electron systems, the iSOI breaks the spin-charge separation (SCS), the hallmark of the Tomonaga-Luttinger liquid~\cite{0034-4885-58-9-002}. As a result, the spin and charge degrees of freedom are intertwined in the collective excitations, which both convey an electric charge and thus both contribute to the system electric response, in contrast to a common case of a purely plasmon-related ballistic conductivity. In addition, the iSOI renormalizes the velocities of the collective excitations. An attractive feature of the iSOI is that the spin-charge structure of the collective excitations in 1D electron systems and their velocities can be tuned by the gate potential.

The iSOI signatures in the dynamics of a 1D electron system were studied in Ref.~\cite{2017arXiv170700316G} in the absence of the RSOI owing to the external electric fields to show that the spin-charge structure of the excitations as well as their velocities can be determined from the Fabry-Pérot resonances in the frequency-dependent conductance of a 1D quantum wire coupled to leads. 

The goal of the present paper is to investigate the interplay of the iSOI and RSOI in the dynamic charge- and spin response of a 1D electron system without contacts that may dramatically affect the system response~\cite{Clarke2016}. 

The search for non-invasive methods to excite the electron system and measure the response is actively pursued nowadays, especially in plasmonics. The tools currently used include the nanoantennas and electron probe techniques~\cite{doi:10.1021/nl200634w}, and even the Kelvin probe force microscopy~\cite{cohen2014observing}.

\begin{figure}[htb]
  \includegraphics[width=0.9\linewidth]{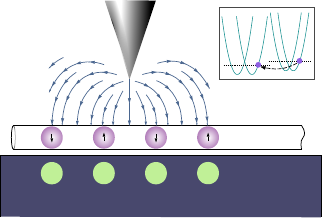}
  \caption{Electrons in a 1D quantum wire with their image charges induced on a gate. Electric field lines from the charged tip acting on the electrons are perpendicular to the wire.}
\label{fig1}
\end{figure}

We consider a single-mode 1D quantum wire subject to an alternating electric field produced by the conducting tip of the scanning probe microscope, as shown in Fig.~\ref{fig1}. Such schematic was discussed in Ref.~\cite{PhysRevB.61.12766} in the context of local disturbance of the charge subsystem~\cite{PhysRevB.57.1515}. We emphasize that the probe electric field, which grows even faster than the potential as the probe approaches the wire, also gives an essential contribution to RSOI thereby disturbing the spin subsystem, too~\cite{PSSR:PSSR201510074}.

The quantum wire is supposed to be placed directly on a conductive gate~\cite{Bachtold1317}, so that the electron image charges on the interface become the source of the iSOI\@. Since the potential difference between the wire and the gate is negligible, the probe electric field screened by the gate has no pulling component along the wire. However, the probe electric field perpendicular to the wire is the source of the time-dependent RSOI\@. 

We show that in response to this, the charge current does appear in the wire in the presence of iSOI and/or RSOI caused by the gate net charge. The RSOI gives rise to an interesting mechanism of electric conductivity. Since the RSOI magnitude is getting modulated along the wire by the non-stationary tip-induced RSOI, there appears a modulation of the bottom of the conduction band that results in the charge current. The process is illustrated by the inset in Fig.~\ref{fig1}. The iSOI produces a complementary conductivity mechanism by mixing the charge- and spin collective excitations, generated by the probe. 

We also find an unusual dependence of the dissipative conductivity on the gate potential. As the potential increases, one out of two collective modes softens, with its amplitude growing. This enhances the current response and the system conductivity as determined from the dissipated power.

\textit{The model}.---We start by formulating the Hamiltonian,
\begin{equation}
\label{fullham}
  H = H_\mathrm{kin} + H_\mathrm{e-e} + H_\mathrm{SOI} + H_\mathrm{ext}\,.
\end{equation}
The kinetic energy is $H_\mathrm{kin} = \sum_{s} \int \psi^+_{s}(x)(\hat{p}_x^2/2m) \psi_{s}(x) dx$, with the electron field operator $\psi_{s}(x)$, the momentum $\hat{p}_{x}$, the spin index $s$.  The $x$ axis is directed along the wire, and $y$ axis is directed normally towards the gate, which is separated by a distance of $a/2$ from the wire. 

The e-e interaction operator reads as
\begin{equation}
\label{Hee}
  \begin{split}
    H_\mathrm{e-e} = &\frac{1}{2} \sum_{s_1 s_2} \int 
     \psi^+_{s_1}(x_1) \psi^+_{s_2}(x_2) U(x_1-x_2) \\ 
    &\! \times \psi_{s_2}(x_2) \psi_{s_1}(x_1)\,dx_1 dx_2\,,
  \end{split}
\end{equation}
where $U(x) = \frac{e^2}{\sqrt{x^2 + d^2}} - \frac{e^2}{\sqrt{x^2 + a^2}}$ is the e-e interaction potential screened by the image charges, $d$ being the quantum wire diameter. Its Fourier transform is $U_q = 2e^2\left[K_0(qd) - K_0(qa)\right]$, with $K_0$ being the modified Bessel function~\cite{olver}. 

A two-particle contribution to the SOI Hamiltonian equals~\cite{PhysRevB.95.045138}
\begin{equation}
\label{isoi}
  \begin{split}
    H_\mathrm{iSOI} = &\frac{\alpha}{2\hbar} \sum_{s_1s_2}  \int \psi^+_{s_1}(x_1) \psi^+_{s_2}(x_2) \left[ E(x_1-x_2)\mathcal{S}_{12} \right. \\
    &{}+ \left. \mathcal{S}_{12} E(x_1-x_2) \right] \psi_{s_2}(x_2) \psi_{s_1}(x_1)\,dx_1 dx_2\,.
  \end{split}
\end{equation} 
Here $\alpha$ is a material-dependent SOI constant, $E(x_i - x_j) = -e a{\left[{(x_i - x_j)}^2 + a^2\right]}^{\! -\frac 32}$ is the $y$ component of the electric field acting on an electron at point $x_2$ from the electron image charge at point $x_1$, and $\mathcal{S}_{12} = (\hat{p}_{x_1} s_1 + \hat{p}_{x_2} s_2)/2$. Eq.~\eqref{isoi} and Eq.~\eqref{Hee} together represent a spin-dependent pair interaction Hamiltonian.

A single-particle contribution to the SOI Hamiltonian comes from the image of the positive background charge density $n_\mathrm{ion}$ in the wire, the charge density $n_\mathrm{g}$ in the gate, and the field of the electron's own image $E(0)$ to give
\begin{equation}
\label{soi0}
  H_\mathrm{RSOI} = \frac{\alpha}{\hbar}\sum_{s}\int \psi^+_{s}(x)\mathfrak{E} \,\hat{p}_{x} s \psi_{s}(x)\, dx\,,
\end{equation}
 with $\mathfrak{E} = E(0) -n_\mathrm{ion}E_0 - 2\pi n_\mathrm{g}$, where $E_0$ is the $q=0$ component of the Fourier-transform $E_q = -2e |q| K_1(|q|a)$ of the field $E(x)$~\cite{olver}. 

Denote the $y$-component of the non-uniform ac-field produced by the probe and screened by the gate by $\mathfrak{F}_{y} = F(x,t)$. Then the external perturbation can be written as
\begin{equation}
\label{hprobe}
	\begin{split}
	  H_\mathrm{ext} &= \frac{\alpha}{2\hbar}\sum_{s}\int dx \,\psi^+_{s}(x) [\mathfrak{F}_{y} \hat{p}_{x} + \hat{p}_{x} \mathfrak{F}_{y} ] s \psi_{s}(x) \\
	  &= - \frac{\alpha m}{e \hbar} \int F(x,t) j_{\sigma}(x)\, dx\,,
	\end{split}
\end{equation}
where $j_{\sigma}(x) = \sum_s s j^{(s)}(x)$ stands for the spin current, with
\begin{equation}
  j^{(s)}(x) = - \frac{i e \hbar}{2 m} \left[ \partial_x \psi^+_s \psi_s(x) - \psi^+_s(x) \partial_x \psi_s \right]\,,
\end{equation}
being the $s$-spin component of the electron current operator.

In order to find a linear response of the system to $H_\mathrm{ext}$ we employ the equation of motion for the Wigner distribution function (WDF) defined as
\begin{equation}
  f^{(s)}(x,p,t) = \frac{1}{2 \pi} \int e^{i p\eta} \left\langle\psi_s^+(x + \frac{\eta}{2},t)\psi_s(x - \frac{\eta}{2},t)\right\rangle \, d\eta \,.
\end{equation}
This technique is particularly well-suited for the problem at hand, since the lack of contacts in the system relieves us from non-trivial problems with the boundary conditions for the WDF~\cite{PhysRevB.88.035401}.

\textit{Results}.---Following Ref.~\cite{PhysRevB.95.045138}, we obtain the following equation for the WDF Fourier transform in the random-phase approximation,
\begin{align}
\label{eqmot}
    &\hbar\omega f^{(s)}_1(q,p,\omega) = \left(\frac{\hbar^2 pq}{m} + \alpha q s E\right) f^{(s)}_1(q,p,\omega) \notag \\
    &{}- \left[f^{(s)}_0(p+\frac{q}{2}) - f^{(s)}_0(p-\frac{q}{2})\right] \times  \\
    &\left \{ \alpha p s F_{q\omega} - \frac{m \alpha}{e \hbar} E_q \sum_{\varsigma}\varsigma j^{(\varsigma)}_{q\omega} + (U_q + \alpha p s E_q) \sum_{\varsigma} n^{(\varsigma)}_{q\omega} \right \} \notag \,.
\end{align}
Here $f^{(s)}_1(q,p,\omega)$ stands for the deviation of $f^{(s)}(q,p,\omega)$ from its equilibrium value $f_0^{(s)}(p)$ as a result of the external perturbation $H_\mathrm{ext}$. Then, $n^{(s)}_{q\omega}$ and $j^{(s)}_{q\omega}$ are, respectively, the electron density and current response, related to the WDF by $n^{(s)}_{q\omega}  = \int f^{(s)}_1(q,p,\omega) dp$ and $j^{(s)}_{q\omega}  = -\frac{e \hbar}{m}\int p f^{(s)}_1(q,p,\omega) dp$. The mean electric field is $E = \mathfrak{E} + n_0 E_0$. The mean electron density $n_0$ is kept fixed, so the Fermi momentum is $k_F^{(s)} = -s k_\mathrm{so} \pm k_F$, where $k_\mathrm{so} = \alpha m E/\hbar^2$ and $k_F$ stands for $\pi n_0/2$.

To derive the closed equations for $n^{(s)}_{q\omega}$, first integrate Eq.~\eqref{eqmot} with respect to $p$:
\begin{align}
\label{conteq}
    \omega e n^{(s)}_{q\omega} + q j^{(s)}_{q\omega} = &\frac{\alpha}{2 \hbar} q s e n_0 F_{q\omega}\\
    &{}+\frac{\alpha q s e}{2\hbar} \left[ 2 E n^{(s)}_{q\omega} + E_q n_0 \sum_{\varsigma} n^{(\varsigma)}_{q\omega}\right]\,.\notag
\end{align}
Substitute $j^{(s)}_{q\omega}$ from Eq.~\eqref{conteq} to Eq.~\eqref{eqmot}, express $f^{(s)}_1(q,p,\omega)$, and integrate the latter with respect to $p$ to get $n^{(s)}_{q\omega}$.

Further notations will be simplified by introducing the dimensionless variables as $\tilde{\alpha} = \dfrac{2}{\pi}\dfrac{\alpha n_0}{e a_B}$, $\mathcal{U}_q = \dfrac{U_q}{\pi \hbar v_F}$, $\mathcal{E} = \dfrac{E}{e n_0^2}$, $\mathcal{E}_q = \dfrac{E_q}{e n_0}$, 
$\mathcal{F}_{q\omega} = \dfrac{F_{q\omega}}{e n_0}$, and $v_q = \dfrac{\omega}{v_F q}$, with $v_F = \dfrac{\hbar k_F}{m}$ and $a_B$ being the Bohr radius in the material. 

The system response to the external perturbation is governed by the following equations ($s = \pm 1$),
\begin{align}
\label{linearsystem}
    &n^{(s)}_{q\omega} \left((\mathcal{E} + \mathcal{E}_q/2 )\mathcal{E}_q \tilde{\alpha}^2 - \mathcal{U}_q + v_q^2 - 1- \tilde{\alpha} s v_q \mathcal{E}_q\right) \notag \\
    &{} + n^{(-s)}_{q\omega}\left( (\mathcal{E} + \mathcal{E}_q/2 )\mathcal{E}_q \tilde{\alpha}^2 - \mathcal{U}_q \right ) = \varphi^{(s)}_{q\omega} \,,
\end{align}
with the spin-dependent perturbation
\begin{equation}
\label{sdp}
  \varphi^{(s)}_{q\omega} = \frac{\tilde{\alpha} s}{2} \mathcal{F}_{q\omega} v_q - \frac{\tilde{\alpha}^2}{2} \mathcal{F}_{q\omega} (\mathcal{E} + \mathcal{E}_q)\,.
\end{equation}

The first term on the right hand side of Eq.~\eqref{sdp} is a perturbation in the spin sector caused directly by the SOI produced by the probe. This term is linear in $\tilde{\alpha}$. The second term describes an indirect perturbation of the charge sector that appears because of the SOI present in the system\@. Its magnitude is, correspondingly, proportional to $\tilde{\alpha}^2$.

The normalized phase velocities of the collective excitations $v=\omega/q v_F$, obtained from Eq.~\eqref{linearsystem} by setting the determinant to zero, are given by
\begin{equation}
\label{dispersion}
    v_{\pm}^2 = 1 + \mathcal{U}_q - \tilde{\alpha}^2\mathcal{E}\mathcal{E}_q
    \pm \sqrt{{\left(\mathcal{U}_q - \tilde{\alpha}^2\mathcal{E}\mathcal{E}_q\right)}^2 + \tilde{\alpha}^2\mathcal{E}_q^2}\,.
\end{equation}

The evolution of the excitation velocities $v_{\pm}$ and the spin-charge separation parameter of the modes that depends on the velocities as $\xi_{\pm} = (v_{\pm} - v_{\pm}^{-1})/\tilde{\alpha}\mathcal{E}_q$ with the change in the iSOI magnitude is analyzed in detail in Refs.~\cite{PhysRevB.95.045138,2017arXiv170700316G}. Here we would like to stress that in the presence of iSOI ($\mathcal{E}_q \ne 0$) both $v_{\pm}$ and $\xi_{\pm}$ can be controlled via the mean electric field $\mathcal{E}$ by tuning the gate potential. Thus, $v_-$ goes to zero as $\mathcal{E}$ grows, i.e.\ the corresponding mode softens. The possibility of tuning the plasmon velocity via the RSOI magnitude was discussed for 2D systems~\cite{doi:10.1063/1.3054645}. An important difference from the 2D case is that without iSOI, $\mathcal{E}$ has no effect on the excitation velocities nor does it violate the SCS between the modes. This is related to the fact that a constant SOI can be completely eliminated in 1D by a unitary transformation~\cite{PhysRevB.88.125143}.

The charge and spin susceptibilities defined by $\chi^{\rho}_{q \omega} = (n^{(+)}_{q\omega} + n^{(-)}_{q\omega})/\mathcal{F}_{q\omega}$ and $\chi^{\sigma}_{q \omega} = (n^{(+)}_{q\omega} - n^{(-)}_{q\omega})/\mathcal{F}_{q\omega}$ are equal to
\begin{equation}
\label{csupt}
  \chi^{\rho}_{q \omega} = \tilde{\alpha}^2 \frac{\mathcal{E}_q + \mathcal{E}(1 - v_q^2)}{(v_q^2 - v_+^2)(v_q^2 - v_-^2)}
\end{equation}
and
\begin{equation}
\label{ssupt}
  \chi^{\sigma}_{q \omega} = \tilde{\alpha} v_q \frac{v_q^2 -1 -2 \mathcal{U}_q + \tilde{\alpha}^2 \mathcal{E}\mathcal{E}_q}{(v_q^2 - v_+^2)(v_q^2 - v_-^2)}\,.
\end{equation}
Their dependence on $\tilde{\alpha}$ is explained similarly to Eq.~\eqref{sdp}.

\begin{figure}[htb]
  \includegraphics[width=0.9\linewidth]{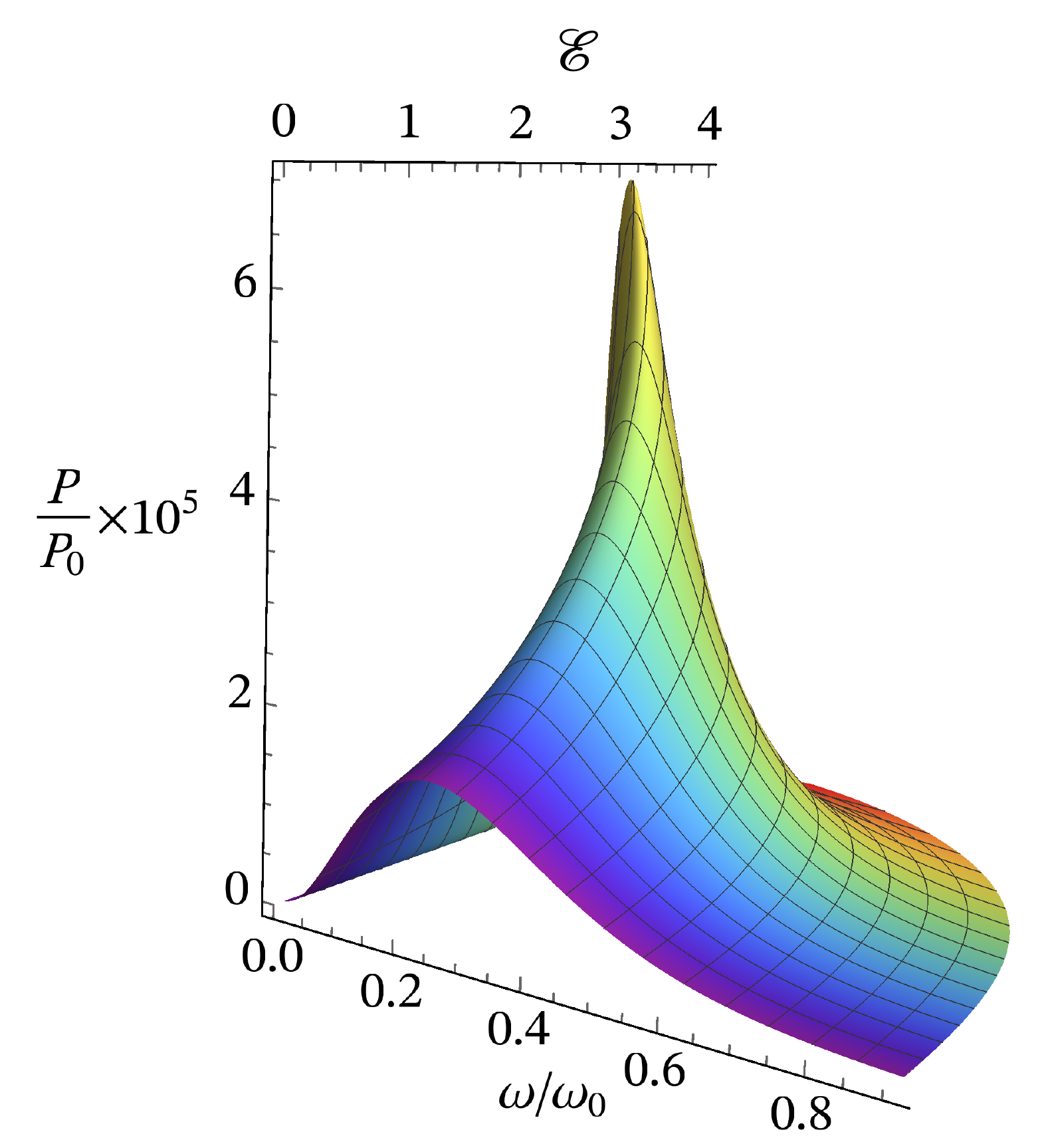}
  \caption{Dissipated power vs.\ perturbation frequency and the external electric field. A tip is modeled by a time-dependent point charge $Q$ lifted by a height $L$ above the wire. Variables are normalized at $\omega_0 = v_F k_F$ and $P_0 = h \omega_0^2 Q^2/e^2$. The system parameters are taken as follows: $k_F a_B = 1.27$, $d = 0.078 a_B$, $a = 0.39 a_B$, $L = 3.9 a_B$, $\tilde{\alpha} = 0.1$.}
\label{fig2}
\end{figure}

According to Eq.~\eqref{hprobe}, the power fed to the system is given by
\begin{equation}
\label{powfed}
  \begin{split}
    P (\omega) &= - \frac{\alpha m}{e \hbar} \int \overline{\frac{\partial F}{\partial t} \langle j_{\sigma}(x) \rangle} \, dx \\
    &= \frac{\tilde{\alpha} \hbar}{4 e} \int_0^{\infty} \omega \mathfrak{Im} \chi^{j_\sigma}_{q \omega} |\mathcal{F}_{q\omega}|^2\, dq\,.
  \end{split}
\end{equation}
The spin current susceptibility $\chi^{j_\sigma}_{q \omega} = \sum_s s j^{(s)}_{q\omega}/\mathcal{F}_{q\omega}$ can be determined from Eq.~\eqref{conteq}, which represents a continuity equation for a system with SOI\@. It is seen that the separate flow of the spin and charge is violated by the second term on the right hand side that refers to inherent mechanisms of mixing the spin and charge degrees of freedom by SOI\@. Using Eq.~\eqref{conteq}, we obtain
\begin{equation}
\label{jsupt}
\chi^{j_\sigma}_{q \omega} = \tilde{\alpha} e v_F \frac{(1 + \tilde{\alpha}^2 \mathcal{E}^2)(1 - v_q^2) + 2 \mathcal{U}_q}{(v_q^2 - v_+^2)(v_q^2 - v_-^2)}\,.
\end{equation}
The imaginary part of the susceptibility for $\omega >0$ equals
\begin{equation}
  \begin{split}
    &\mathfrak{Im} \chi^{j_\sigma}_{q \omega} = \left[\frac{(1 - v_{-}^2)(1 + \tilde{\alpha}^2 \mathcal{E}^2) + 2 \mathcal{U}_q}{2 v_{-}(v_{+}^2 - v_{-}^2)} \delta (\omega - q v_{-} v_F)\right.\\
    &\left. + \frac{(v_{+}^2 - 1)(1 + \tilde{\alpha}^2 \mathcal{E}^2) + 2 \mathcal{U}_q}{2 v_{+}(v_{+}^2 - v_{-}^2)} \delta (\omega - q v_{+} v_F) \right]\pi \tilde{\alpha} e v_F^2 q\,.
  \end{split}
\end{equation}
The leading contribution to the dissipated power comes from the first $\delta$-function,
\begin{equation}
  P (\omega) = \tilde{\alpha}^2 h \omega^2 |\mathcal{F}_{q\omega}|^2 \frac{(1 - v_{-}^2)(1 + \tilde{\alpha}^2 \mathcal{E}^2) + 2 \mathcal{U}_q}{16 v_{-}^3(v_{+}^2 - v_{-}^2)(1 + \frac{\omega}{v_{-}^2 v_F}\frac{\partial v_{-}}{\partial q})}\,,
\end{equation}
with $q$ determined from $\omega = q v_{-}(q) v_F$.
The dependence of the excitation velocity $v_{-}(\mathcal{E})$ on the electric field of the gate results in a sharp peak in $P (\omega,\mathcal{E})$, as illustrated by Fig.~\ref{fig2}. 

The dissipated heat could be measured by the scanning thermal microscopy~\cite{Lee2013}, but a detailed consideration of the heat release involving the kinetics of the phonon subsystem is beyond the scope of the present letter.

\textit{Conclusion}.---To summarize, the dynamic charge and spin response of a 1D electron system to an alternating electric field of the charged probe was investigated in the presence of the SOI\@. The electric response to the probe-induced non-stationary SOI appears because of the RSOI and iSOI present in the system. As a result of the interplay between the iSOI and RSOI, the velocities of the collective excitations and their spin-charge structure become tunable via the electric field of the gate, and so does the system conductivity determined from the dissipated power.

I am grateful to Vladimir Sablikov for helpful discussions. This work was partially supported by Russian Foundation for Basic Research (Grant No 17--02--00309) and Russian Academy of Sciences.

\bibliography{paper}

\end{document}